\documentclass[12pt]{article}
\usepackage{makeidx}
\usepackage{amsmath}
\usepackage{amsfonts}
\usepackage{amssymb}
\usepackage{epsfig}
\usepackage[cp1251]{inputenc}
\usepackage{graphicx}

\newcounter{defin}
\newcounter{lemma}
\newcounter{theorem}
\newcounter{proposition}
\newcounter{example}
\newenvironment{lemma}{\par\refstepcounter{lemma}     \textbf{Lemma \thelemma.} }{\rm\par}
\newenvironment{theorem}{\par\refstepcounter{theorem}     \textbf{Theorem \thetheorem.}\ }{\rm\par}

\begin{document}

\title{On the classical capacity of quantum Gaussian measurement}
\author{A. S. Holevo \\
Steklov Mathematical Institute, RAS, Moscow, Russia}
\date{\today}
\maketitle

\begin{abstract}
In this paper we consider the classical capacity problem for Gaussian
measurement channels without imposing any kind of threshold condition. We
prove Gaussianity of the average state of the optimal ensemble in general
and discuss the Hypothesis of Gaussian Maximizers concerning the structure
of the ensemble. The proof uses an approach of Wolf, Giedke and Cirac
adapted to the convex closure of the output differential entropy. Then we
discuss the case of one mode in detail, including the dual problem of
accessible information of a Gaussian ensemble.

In quantum communications there are several studies of the classical
capacity in the transmission scheme where not only the Gaussian channel but
also the receiver is fixed, and the optimization is performed over certain
set of the input ensembles. These studies are practically important in view
of the complexity of the optimal receiver in the Quantum Channel Coding
(HSW) theorem. Our findings are relevant to such a situation where the
receiver is Gaussian and concatenation of the channel and the receiver can
be considered as one Gaussian measurement channel. Our efforts in this and
preceding papers are then aimed at establishing full Gaussianity of the
optimal ensemble (usually taken as an assumption) in such schemes.
\end{abstract}

\section{Introduction}



From the viewpoint of information theory measurements are hybrid
communication channels that transform input quantum states into classical
output data. As such, they are described by the classical information
capacity which is the most fundamental quantity characterizing their
ultimate information-processing performance \cite{hall}, \cite{da}, \cite%
{Oreshkov}, \cite{h5}. Channels with continuous output, such as bosonic
Gaussian measurements, do not admit direct embedding into properly quantum
channels and hence require separate treatment. In particular, their output
entropy is the Shannon differential entropy, instead of the quantum entropy,
which completely changes the pattern of the capacity formulas. The classical
capacity of multimode Gaussian measurement channels was computed in \cite%
{acc-noJ} under so called threshold condition (which includes
phase-insensitive or gauge covariant channels as a special case). The
essence of this condition is that it reduces the classical capacity problem
to the minimum output differential entropy problem solved in \cite{ghm} (in
the context of quantum Gaussian channels a similar condition was introduced
and studied in \cite{scha}, \cite{h2}, see also references therein).

In this paper we consider the classical capacity problem for Gaussian
measurement channels without imposing any kind of threshold condition. In
particular, in the framework of quantum communication, this means that both
(noisy) heterodyne and (noisy/noiseless) homodyne measurements \cite{caves}
are treated from a common viewpoint. In this setting, we prove Gaussianity
of the average state of the optimal ensemble in general and discuss the
Hypothesis of Gaussian Maximizers (HGM) concerning the structure of the
ensemble. The proof uses the approach of the paper of Wolf, Giedke and Cirac
\cite{wgc} applied to the convex closure of the output differential entropy.
Then we discuss the case of one mode in detail, including the dual problem
of accessible information of a Gaussian ensemble.

In quantum communications there are several studies of the classical
capacity in the transmission scheme where not only the Gaussian channel but
also the receiver is fixed, and the optimization is performed over certain
set of the input ensembles (see \cite{caves}, \cite{hall3}, \cite{guha},
\cite{lee} and references therein). These studies are practically important
in view of the enormous complexity of the optimal receiver in the Quantum
Channel Coding (HSW) theorem (see e.g. \cite{QSCI}). Our findings are
relevant to such a situation where the receiver is Gaussian and
concatenation of the channel and the receiver can be considered as one
Gaussian measurement channel. Our efforts in this and preceding papers are
then aimed at establishing full Gaussianity of the optimal ensemble (usually
taken as a key assumption) in such schemes.

\section{The measurement channel and its classical capacity}

An \textit{ensemble} $\mathcal{E}=\left\{ \pi (dx),\rho (x)\right\} $
consists of probability measure $\pi (dx)$ on a standard measurable space $%
\mathcal{X}$ and a measurable family of density operators (quantum states) $%
x\rightarrow \rho (x)$ on the Hilbert space $\mathcal{H}$ of the quantum
system. The \textit{average state} of the ensemble is the barycenter of this
measure%
\begin{equation*}
\bar{\rho}_{\mathcal{E}}=\int_{\mathcal{X}}\rho (x)\,\pi (dx),
\end{equation*}%
the integral existing in the strong sense in the Banach space of trace-class
operators on $\mathcal{H}$.

Let $M=\{M(dy)\}$ be an observable (POVM) on $\mathcal{H}$ with the outcome
standard measurable space $\mathcal{Y}$. There exists a $\sigma -$finite
measure $\mu (dy)$ such that for any density operator $\rho $ the
probability measure $\mathrm{Tr}\rho M(dy)$ is absolutely continuous w.r.t. $%
\mu (dy),$ thus having the probability density $p_{\rho }(y)$ (one can take $%
\mu (dy)=\mathrm{Tr}\rho _{0}M(dy)$ where $\rho _{0}$ is a nondegenerate
density operator). The affine map $M:\rho \rightarrow p_{\rho }(\cdot )$
will be called the \textit{measurement channel}.

The joint probability distribution of $x,y$ on $\mathcal{X\times Y}$ \ is
uniquely defined by the relation%
\begin{equation*}
P(A\times B)=\int_{A}\pi (dx)\mathrm{Tr}\,\rho (x)M(B)=\mathrm{Tr}%
\int_{A}\int_{B}\,p_{\rho (x)}(y)\,\pi (dx)\mu (dy),
\end{equation*}%
where $A$ is an arbitrary Borel subset of $\mathcal{X}$ and $B$ is that of $%
\mathcal{Y}.$ The classical Shannon information between $x,y$ is equal to%
\begin{equation*}
I(\mathcal{E},M)=\int \int \pi (dx)\mu (dy)p_{\rho (x)}(y)\log \frac{p_{\rho
(x)}(y)}{p_{\bar{\rho}_{\mathcal{E}}}(y)}
\end{equation*}%
In what folows we will consider POVMs having (uniformly) bounded operator
density, $M(dy)=m(y)\mu (dy),$ with $\left\Vert m(y)\right\Vert \leq b,$ so
that the probability densities $p_{\rho }(y)=\mathrm{Tr}\,\rho m(y)$ are
uniformly bounded, $0\leq p_{\rho }(y)\leq b$. (The probability densities
corresponding to Gaussian observables we will be dealing with possess this
property). Moreover, without loss of generality \cite{acc}\ we can assume $%
b=1.$ Then the output differential entropy
\begin{equation}
h_{M}(\rho )=-\int p_{\rho }(y)\log \,p_{\rho }(y)\mu (dy)  \label{den}
\end{equation}%
is well defined with values in $[0,+\infty ]$ (see \cite{acc} for the
detail). The output differential entropy is concave lower semicontinuous
(w.r.t. trace norm) functional of a density operator $\rho $. The concavity
follows from the fact that the function $p\rightarrow -p\log p,\,p\in $ $%
[0,1]$ is convave. Lower semicontinuity follows by an application of the
Fatou-Lebesgue lemma from the fact that this function is nonnegative,
continuous and $\left\vert p_{\rho }(y)-p_{\sigma }(y)\right\vert \leq
\left\Vert \rho -\sigma \right\Vert _{1}.$

Next we define the \textit{convex closure of the output differential entropy}
(\ref{den}):
\begin{equation}
e_{M}(\rho )=\inf_{\mathcal{E}:\bar{\rho}_{\mathcal{E}}=\rho }\int
h_{M}(\rho (x))\pi (dx),  \label{e1}
\end{equation}%
which is the \textquotedblleft measurement channel analog\textquotedblright\
of the convex closure of the output entropy for a quantum channel \cite{Shir}%
.

\begin{lemma}
\label{l1} \textit{The functional $e_{M}(\rho )$ is convex, lower
semicontinuous and strongly superadditive:%
\begin{equation}
e_{M_{1}\otimes M_{2}}(\rho _{12})\geq e_{M_{1}}(\rho _{1})+e_{M_{2}}(\rho
_{2}).  \label{ssa}
\end{equation}%
}
\end{lemma}

As it is well known, the property (\ref{ssa}) along with the definition (\ref%
{e1}) imply \textit{additivity}: if $\rho _{12}=\rho _{1}\otimes \rho _{2}$
then%
\begin{equation}
e_{M_{1}\otimes M_{2}}(\rho _{12}) = e_{M_{1}}(\rho _{1})+e_{M_{2}}(\rho
_{2}).  \label{adit}
\end{equation}

\textit{Proof.} The lower semicontinuity follows from the similar property
of the output differential entropy much in the same way as in the case of
quantum channels, treated in \cite{Shir}, Proposition 4, see also \cite%
{Shir1}, Proposition 1.

Let us prove strong superadditivity. Let
\begin{equation}
\rho _{12}=\int \rho _{12}(x)\pi (dx)  \label{decom}
\end{equation}%
be a decomposition of a density operator $\rho _{12}$ on $\mathcal{H}%
_{1}\otimes \mathcal{H}_{2}$, then
\begin{eqnarray*}
&&p_{M_{1}\otimes M_{2}}(y_{1},y_{2}|x) \\
&=&\mathrm{Tr}\,\rho _{12}(x)\left[ m_{1}(y_{1})\otimes m_{2}(y_{2})\right]
\\
&=&\mathrm{Tr}\,\rho _{1}(x)\,m_{1}(y_{1})\,\mathrm{Tr}\,\rho
_{2}(y_{1},x)\,m_{2}(y_{2}) \\
&=&p_{M_{1}}(y_{1}|x)\,p_{M_{2}}(y_{2}|y_{1},x),
\end{eqnarray*}%
where $\,\rho _{1}(x)=\mathrm{Tr}_{2}\,\rho _{12}(x),\rho _{2}(y_{1},x)=%
\frac{\mathrm{Tr}_{1}\,\rho _{12}(x)\left[ m_{1}(y_{1})\otimes I_{2}\right]
}{\mathrm{Tr}\,\rho _{12}(x)\left[ m_{1}(y_{1})\otimes I_{2}\right] },$ so
that
\begin{equation*}
\mathrm{Tr}\,\rho _{12}(x)\left[ m_{1}(y_{1})\otimes I_{2}\right] =\mathrm{Tr%
}\,\rho _{1}(x)\,m_{1}(y_{1})=p_{M_{1}}(y_{1}|x),
\end{equation*}%
and $\rho _{2}=\int \int \rho _{2}(y_{1},x)p_{M_{1}}(y_{1}|x)\pi (dx)\mu
_{1}(dy_{1})$ while $\rho _{1}=\int \rho _{1}(x)\pi (dx).$ It follows
\begin{eqnarray*}
h(Y_{1},Y_{2}|X) &\equiv &\int h_{M_{1}\otimes M_{2}}(\rho _{12}(x))\pi (dx)
\\
&=&\int h_{M_{1}}(\rho _{1}(x))\pi (dx) \\
&+&\int \int h_{M_{2}}(\rho _{2}(y_{1},x))p_{M_{1}}(y_{1}|x)\pi (dx)\mu
_{1}(dy_{1}) \\
&=&h(Y_{1}|X)+h(Y_{2}|Y_{1},X),
\end{eqnarray*}%
whence taking the infimum over decompositions (\ref{decom}), we obtain (\ref%
{ssa}). $\square $

Let $H$ be a Hamiltonian in the Hilbert space $\mathcal{H}$ of the quantum
system, $E$ a positive number. Then the \textit{energy-constrained classical
capacity} of the channel $M$ is equal to
\begin{equation}
C(M,H,E)=\sup_{\mathcal{E}:\mathrm{Tr}\bar{\rho}_{\mathcal{E}}H\leq E}I(%
\mathcal{E},M),  \label{0}
\end{equation}%
where maximization is over the input ensembles of states $\mathcal{E}$
satisfying the energy constraint $\mathrm{Tr}\bar{\rho}_{\mathcal{E}}H\leq E$%
, as shown in \cite{acc-noJ}, proposition 1.

If $h_{M}(\bar{\rho}_{\mathcal{E}})<+\infty $, then
\begin{equation}
I(\mathcal{E},M)=h_{M}(\bar{\rho}_{\mathcal{E}})-\int h_{M}(\rho (x))\pi
(dx).  \label{i1}
\end{equation}%
Note that the measurement channel is entanglement-breaking \cite{QSCI} hence
its classical capacity is additive and is given by the one-shot expression (%
\ref{0}). By using (\ref{i1}), (\ref{e1}), we obtain%
\begin{equation}
C(M,H,E)=\sup_{\rho :\mathrm{Tr}\rho H\leq E}\left[ h_{M}(\rho )-e_{M}(\rho )%
\right] .  \label{C1}
\end{equation}

\section{Gaussian maximizers for multimode bosonic Gaussian observable}

Consider now multimode bosonic Gaussian system with the quadratic
Hamiltonian $H=R\epsilon R^{t},$ where $\epsilon >0$ is the energy matrix,
and $R=\left[ q_{1},p_{1},\dots ,q_{s},p_{s}\right] $ is the row vector of
the bosonic position-momentum observables, satisfying the canonical
commutation relation
\begin{equation*}
\lbrack R^{t},R]=i\Delta I,\quad \Delta =\mathrm{diag}\left[
\begin{array}{cc}
0 & 1 \\
-1 & 0%
\end{array}%
\right] _{\overline{1,\dots ,s}},
\end{equation*}%
(see e.g. \cite{QSCI}, \cite{sera}). \textit{From now on we will consider
only states with finite second moments}. For such states $h_{M}(\rho )\leq
h_{M}(\rho _{\alpha })<+\infty ,$ where ${\alpha }$ is the covariance matrix
of $\rho ,$ by the maximum entropy principle. For \textit{centered} states
(i.e. states with vanishing first moments) the covariance matrix and the
matrix of second moments coincide and are equal to
\begin{equation*}
\alpha =\mathrm{Re}\,\mathrm{Tr}R^{t}\rho R.
\end{equation*}%
%
%
%
%
%
%
The energy constraint reduces to \footnote{%
We denote Sp trace of $s\times s$-matrices as distinct from trace of
operators on $\mathcal{H}$.}
\begin{equation}
\mathrm{Sp}\,\alpha \,\epsilon \,\leq E.  \label{E1}
\end{equation}

We denote the set of all states $\rho $ with the fixed covariance matrix $%
\alpha $ by $\mathfrak{S}(\alpha )$ and we will study the following $\alpha $%
-\textit{constrained} capacity
\begin{equation}
C(M;\alpha )=\sup_{\mathcal{E}:\bar{\rho}_{\mathcal{E}}\in \mathfrak{S}%
(\alpha )}I(\mathcal{E},M)=\sup_{\rho \in \mathfrak{S}(\alpha )}\left[
h_{M}(\rho )-e_{M}(\rho )\right] .  \label{ca}
\end{equation}%
With the Hamiltonian $H=R\epsilon R^{t},$ the \textit{energy-constrained
classical capacity} of observable $M$ is
\begin{equation*}
C(M;H,E)=\sup_{\alpha :\mathrm{Sp}\,\alpha \epsilon \leq E}C(M;\alpha ).
\end{equation*}

We will be interested in the approximate position-momentum measurement
(observable, POVM)
\begin{equation}
M(d^{2s}z)=D(z)\rho _{\beta }D(z)^{\ast }\frac{d^{2s}z}{\left( 2\pi \right)
^{s}}  \label{MTBs}
\end{equation}%
where $\rho _{\beta }$ is centered Gaussian density operator with the
covariance matrix $\beta $ and
\begin{equation*}
D(z)=\exp i\sum_{j=1}^{s}\left( y_{j}q_{j}-x_{j}p_{j}\right) ,\quad z=\left[
\begin{array}{ccc}
x_{1},y_{1}, & \dots , & x_{s},y_{s}%
\end{array}%
\right] ^{t}\in \mathbb{R}^{2s}
\end{equation*}%
are the unitary displacement operators. Thus $\mu (dz)=\frac{d^{2s}z}{\left(
2\pi \right) ^{s}}$ and the operator-valued density of POVM (\ref{MTBs}) is $%
m(z)=D(z)\rho _{\beta }D(z)^{\ast }.$

In what follows we will consider $n$ independent copies of our bosonic
system on the Hilbert space $\mathcal{H}^{\otimes n}.$ We will supply all
the quantities related to $k-$th copy ($k=1,\dots ,n$) with upper index $%
^{(k)}$, and we will use tilde to denote quantities related to the whole
collection on $n$ copies. Thus%
\begin{equation*}
\tilde{z}=\left[
\begin{array}{c}
z^{(1)} \\
\dots \\
z^{(n)}%
\end{array}%
\right] ,\quad D(\tilde{z})=D(z^{(1)})\otimes \dots \otimes D(z^{(n)})
\end{equation*}%
and
\begin{equation*}
M^{\otimes n}(d\tilde{z})=\tilde{m}(\tilde{z})\tilde{\mu}(d\tilde{z})=\left[
m(z^{(1)})\otimes \dots \otimes m(z^{(n)})\right] \,\mu (dz^{(1)})\dots \mu
(dz^{(n)}).
\end{equation*}

\begin{lemma}
\label{l2} \textit{Let $O=\left[ O_{kl}\right] _{k,l=1,\dots ,n}$ be a real
orthogonal $n\times n-$matrix and $U$ -- the unitary operator on $\mathcal{H}%
^{\otimes n}$ corresponding to the linear symplectic transformation%
\begin{equation*}
\tilde{R}=\left[
\begin{array}{ccc}
R^{(1)}, & \dots , & R^{(n)}%
\end{array}%
\right] \rightarrow \tilde{R}\,O,\quad
\end{equation*}%
so that
\begin{equation}
U^{\ast }D(\tilde{z})U=D(O\,\tilde{z}).  \label{udo}
\end{equation}
Then for any state $\tilde{\rho}$ on $\mathcal{H}^{\otimes n}$
\begin{equation}
e_{M^{\otimes n}}(\tilde{\rho})=e_{M^{\otimes n}}(U\tilde{\rho}U^{\ast }).
\label{emr}
\end{equation}%
}
\end{lemma}

\textit{Proof.} The covariance matrix $\tilde{\beta}$ of $\rho _{\beta
}^{\otimes n}$ is block-diagonal, $\tilde{\beta}=[\delta _{kl}\beta
]_{k,l=1,\dots ,n}$, hence $O^{t}\tilde{\beta}O=\tilde{\beta}$. Thus we have
$U^{\ast }\rho _{\beta }^{\otimes n}U=\rho _{\beta }^{\otimes n},$ and
taking into account (\ref{udo}),
\begin{equation*}
U^{\ast }\tilde{m}(\tilde{z})U=D(O\,\tilde{z})\rho _{\beta }^{\otimes n}D(O\,%
\tilde{z})^{\ast }=\tilde{m}(O\tilde{z}).
\end{equation*}%
Therefore for any state $\tilde{\sigma}$ on $\mathcal{H}^{\otimes n}$ the
output probability density of the measurement channel $\tilde{M}=M^{\otimes
n}$ corresponding to the input state $U\tilde{\sigma}U^{\ast }$ is
\begin{equation}
p_{U\tilde{\sigma}U^{\ast }}(\tilde{z})=\mathrm{Tr}\,\left( U\tilde{\sigma}%
U^{\ast }\right) \tilde{m}(\tilde{z})=\mathrm{Tr}\,\tilde{\sigma}\tilde{m}(O%
\tilde{z})=p_{\tilde{\sigma}}(O\tilde{z}).  \label{pinv}
\end{equation}%
Hence, by using orthogonal invariance of the Lebesgue measure,
\begin{equation*}
h_{M^{\otimes n}}(U\tilde{\sigma}U^{\ast })=h_{M^{\otimes n}}(\tilde{\sigma}%
).
\end{equation*}

If $\tilde{\rho}=\int_{\mathcal{X}}\tilde{\rho}(x)\,\pi (dx),$ then $U\tilde{%
\rho}U^{\ast }=\int_{\mathcal{X}}\left( U\tilde{\rho}(x)\,U^{\ast }\right)
\pi (dx),$ and taking $\tilde{\sigma}=\tilde{\rho}(x)$ in the previous
formula, we deduce
\begin{equation*}
\int_{\mathcal{X}}h_{M^{\otimes n}}(U\tilde{\rho}(x)U^{\ast })\pi (dx)=\int_{%
\mathcal{X}}h_{M^{\otimes n}}(\tilde{\rho}(x))\pi (dx),
\end{equation*}%
hence (\ref{emr}) follows. $\square $

\begin{lemma}
\label{l3} \textit{Let $M$ be the Gaussian measurement (\ref{MTBs}). For any
state $\rho $ with finite second moments $e_{M}(\rho )\geq e_{M}(\rho
_{\alpha })$ where ${\alpha }$ is the covariance matrix of $\rho $.}
\end{lemma}

\textit{Proof.} The proof follows the pattern of Lemma 1 from the paper of
Wolf, Giedke and Cirac \cite{wgc}. Without loss of generality we can assume
that $\rho $ is centered. We have%
\begin{equation}
e_{M}(\rho )\overset{(1)}{=}\frac{1}{n}e_{M^{\otimes n}}(\rho ^{\otimes n})%
\overset{(2)}{=}\frac{1}{n}e_{M^{\otimes n}}(\tilde{\rho})\overset{(3)}{\geq
}\frac{1}{n}\sum_{k=1}^{n}e_{M}(\tilde{\rho}^{(k)}),  \label{chain}
\end{equation}%
where $\tilde{\rho}=U\rho ^{\otimes n}U^{\ast }$ with symplectic unitary $U$
in $\mathcal{H}^{\otimes n},\,$ corresponding to an orthogonal matrix $O$ as
in lemma \ref{l2}, and $\tilde{\rho}^{(k)}$ is the $k-$th partial state of $%
\tilde{\rho}.$

Step (1) follows from the additivity (\ref{adit}). Step (2) follows from
lemma \ref{l2}, and step (3) follows from the superadditivity of $e_{M}$
(lemma \ref{l1}). The final step of the proof
\begin{equation}
\liminf_{n\to\infty}\frac{1}{n}\sum_{k=1}^{n}e_{M}(\tilde{\rho}^{(k)})\geq
e_{M}(\rho _{\alpha })  \label{s4}
\end{equation}%
uses ingeniously constructed $U$ from \cite{wgc} and lower semicontinuity of
$e_{M}$ (lemma \ref{l1}). Namely, $n=2^{m},$ and $U$ corresponds via (\ref%
{udo}) to the following special orthogonal matrix
\begin{equation*}
O=\left[ O_{kl}\right] _{k,l=1,\dots ,n}=H^{\otimes m},\quad H=\frac{1}{%
\sqrt{2}}\left[
\begin{array}{cc}
1 & 1 \\
1 & -1%
\end{array}%
\right] .
\end{equation*}%
Every row of the $n\times n-$matrix $O$ except the first one which has all
the elements 1, has $n/2=2^{m-1}$ elements equal to 1 and $n/2$ elements
equal to -1. Then the quantum characteristic function of the states $\tilde{%
\rho}^{(k)},k=2,\dots ,n$ is equal to $\phi (z/\sqrt{n})^{n/2}\phi (-z/\sqrt{%
n})^{n/2}$, where $\phi (z)$ is the quantum characteristic function of the
state $\rho .$ This allows to apply Quantum Central Limit Theorem \cite{huds}
to show that $\tilde{\rho}^{(k)}\rightarrow \rho _{\alpha }$ as $%
n\rightarrow \infty ,$ in a uniform way, implying (\ref{s4}), see \cite{wgc}
for detail. $\square $

\begin{theorem}
\label{t1} \textit{The optimizing density operator $\rho $ in (\ref{ca}) is
the (centered) Gaussian density operator $\rho _{\alpha }:$%
\begin{equation}
C(M;\alpha )=h_{M}(\rho _{\alpha })-e_{M}(\rho _{\alpha }),  \label{cma}
\end{equation}%
and hence%
\begin{equation}
C(M,H,E)=\max_{\alpha :\mathrm{Sp}\,\alpha \,\epsilon \,\leq E}C(M;\alpha
)=\max_{\alpha :\mathrm{Sp}\,\alpha \,\epsilon \,\leq E}\left[ h_{M}(\rho
_{\alpha })-e_{M}(\rho _{\alpha })\right] .  \label{cmhe}
\end{equation}%
}
\end{theorem}

\textit{Proof.} Lemma \ref{l3} implies that for any $\rho $ with finite
second moments $e_{M}(\rho )\geq e_{M}(\rho _{\alpha })$ where ${\alpha }$
is the covariance matrix of $\rho $. On the other hand, by the maximum
entropy principle, $h_{M}(\rho )\leq h_{M}(\rho _{\alpha })$. Hence (\ref%
{cma}) is maximized by a Gaussian density operator. $\square $

\textbf{Remark.} The proof of lemma \ref{l2} and hence of theorem \ref{t1}
can be extended to a general Gaussian observable $M$ in the sense of \cite%
{QSCI}, \cite{hclass}, defined via operator-valued characteristic function
of the form%
\begin{equation}
\phi _{M}(w)=\exp \left( i\,R\,Kw-\frac{1}{2}w^{t}\beta w\right) ,
\label{cf}
\end{equation}%
where $K$ is a scaling matrix, $\beta \geq \pm \frac{i}{2}K^{t}\Delta K$, by
using this function to obtain generalization of the relation (\ref{pinv})
for the measurement probability densities. The case (\ref{MTBs}) corresponds
to the type 1 Gaussian observable with $K=I_{2s}$.

\textbf{Hypothesis of Gaussian Maximizers (HGM)}: \textit{Let $M$ be an
arbitrary Gaussian observable. Then the optimal ensemble for (\ref{e1}) and
hence for (\ref{0}) is Gaussian, more precisely it consists of (properly
squeezed) coherent states with the displacement parameter having Gaussian
probability distribution.}

For Gaussian measurement channels of the type 1 (essentially of the form (%
\ref{MTBs}), see \cite{hclass} for complete classification) and Gaussian
states $\rho _{\alpha }$ satisfying the \textquotedblleft threshold
condition\textquotedblright\ we have
\begin{equation}
e_{M}(\rho _{\alpha })=\min_{\rho }h_{M}(\rho ),  \label{e=min}
\end{equation}%
with the minimum attained on a squeezed coherent state, which implies the
validity of the HGM and an efficient computation of $C(M,H,E)$, see \cite%
{acc-noJ}. On the other hand, the problem remains open in the case where the
\textquotedblleft threshold condition\textquotedblright\ is violated, and in
particular, for all Gaussian measurement channels of the type 2, with the
generic example of the energy-constrained approximate measurement of the
position $\left[ q_{1},\dots ,q_{s}\right] $ subject to Gaussian noise (see
\cite{hy}, where the entanglement-assisted capacity of such a measurement
was computed). In the following section we discuss in some detail the HGM in
this case for one mode system.

\section{Gaussian measurements in one mode}

Our framework in this section will be one bosonic mode described by the
canonical position and momentum operators $q,\,p$ . We recall that
\begin{equation*}
D(x,y)=\exp i\left(yq-xp\right),\quad x,y\in \mathbb{R}
\end{equation*}%
are the unitary displacement operators.

We will be interested in the obserbable
\begin{equation}
M(dxdy)=D(x,y)\rho _{\beta }D(x,y)^{\ast }\frac{dxdy}{2\pi },  \label{MTB}
\end{equation}%
where $\rho _{\beta }$ is centered Gaussian density operator with the
covariance matrix
\begin{equation}
\beta =\left[
\begin{array}{cc}
\beta _{q} & 0 \\
0 & \beta _{p}%
\end{array}%
\right] ;\quad \beta _{q}\beta _{p}\geq \frac{1}{4}.  \label{beta}
\end{equation}

Let $\rho _{\alpha }$ be a centered Gaussian density operator with the
covariance matrix%
\begin{equation}
\alpha =\left[
\begin{array}{cc}
\alpha _{q} & 0 \\
0 & \alpha _{p}%
\end{array}%
\right] .  \label{alpha}
\end{equation}%
The problem is to compute $e_{M}(\rho _{\alpha })$ and hence the classical
capacity $C(M,H,E)$ for the oscillator Hamiltonian $H=\frac{1}{2}\left(
q^{2}+p^{2}\right) $ (as shown in the Appendix of \cite{hy}, we can restrict
to Gaussian states $\rho _{\alpha }$ with the diagonal covariance matrix in
this case). The energy constraint (\ref{E1}) takes the form%
\begin{equation}
\alpha _{q}+\alpha _{p}\leq 2E.  \label{E11}
\end{equation}

The measurement channel corresponding to POVM (\ref{MTB}) acts on the
centered Gaussian state $\rho _{\alpha }\ $ by the formula
\begin{eqnarray}
M &:&\rho _{\alpha }\rightarrow p_{\rho _{\alpha }}(x,y)  \label{pdmtb} \\
&=&\frac{1}{\sqrt{2\pi \left( \alpha _{q}+\beta _{q}\right) \left( \alpha
_{p}+\beta _{p}\right) }}\exp \left[ -\frac{x^{2}}{2\left( \alpha _{q}+\beta
_{q}\right) }-\frac{y^{2}}{2\left( \alpha _{p}+\beta _{p}\right) }\right] ,
\notag
\end{eqnarray}%
so that \footnote{%
In this expression $c$ is a fixed constant depending on the normalization of
the underlying measure $\mu$ in (\ref{den}). It does not enter the
information quantities which are differences of the two differential
entropies.}%
\begin{equation}
h_{M}(\rho _{\alpha })=\frac{1}{2}\log \left( \alpha _{q}+\beta _{q}\right)
\left( \alpha _{p}+\beta _{p}\right)+c.  \label{ha}
\end{equation}

\textit{Assuming validity of the HGM}, we will optimize over ensembles of
squeezed coherent states
\begin{equation*}
\rho _{x,y}=D(x,y)\,\rho _{\Lambda }D(x,y)^{\ast },\quad (x,y)\in \mathbb{R}%
^{2},
\end{equation*}%
where $\rho _{\Lambda }$ is centered Gaussian state with correlation matrix $%
\Lambda =\left[
\begin{array}{cc}
\delta & 0 \\
0 & 1/\left( 4\delta \right)%
\end{array}%
\right] ,$ and the vector $(x,y)$ has centered Gaussian distribution with
covariance matrix $\left[
\begin{array}{cc}
\gamma _{q} & 0 \\
0 & \gamma _{p}%
\end{array}%
\right] .$ Then the average state $\bar{\rho}_{\mathcal{E}}$ of the ensemble
is centered Gaussian $\rho _{\alpha }$ with the covariance matrix (\ref%
{alpha}), where
\begin{equation*}
\alpha _{q}=\gamma _{q}+\delta ,\quad \alpha _{p}=\gamma _{p}+1/\left(
4\delta \right) ,
\end{equation*}%
hence%
\begin{equation}
\frac{1}{4\alpha _{p}}\leq \delta \leq \alpha _{q}.  \label{inter}
\end{equation}%
For this ensemble%
\begin{equation*}
\int h_{M}(\rho _{x,y})\pi (dx\,dy)=h_{M}(\rho _{\Lambda })=\frac{1}{2}\log
\left( \delta +\beta _{q}\right) \left( 1/\left( 4\delta \right) +\beta
_{p}\right) +c.
\end{equation*}%
Then the hypothetical value
\begin{equation}
e_{M}(\rho _{\alpha })=\min_{1/\left( 4\alpha _{p}\right) \leq \delta \leq
\alpha _{q}}\frac{1}{2}\log \left( \delta +\beta _{q}\right) \left( 1/\left(
4\delta \right) +\beta _{p}\right) +c.  \label{em}
\end{equation}%
The derivative of the minimized expression vanishes for $\delta =\frac{1}{2}%
\sqrt{\frac{\beta _{q}}{\beta _{p}}}.$ Thus, depending on the position of
this value with respect to the interval (\ref{inter}), we obtain three
possibilities:%
\begin{equation*}
\begin{tabular}{|c|c|c|c|}
\hline
\multicolumn{4}{|c|}{$\text{Table 1 }$} \\ \hline
range & L: $\frac{1}{2}\sqrt{\frac{\beta _{q}}{\beta _{p}}}<\frac{1}{4\alpha
_{p}}$ & C:$\frac{1}{4\alpha _{p}}\leq \frac{1}{2}\sqrt{\frac{\beta _{q}}{%
\beta _{p}}}\leq \alpha _{q}$ & R: $\alpha _{q}<\frac{1}{2}\sqrt{\frac{\beta
_{q}}{\beta _{p}}}$ \\ \hline
HGM & open & valid & open \\ \hline
$\delta _{opt}$ & $1/\left( 4\alpha _{p}\right) $ & $\frac{1}{2}\sqrt{\frac{%
\beta _{q}}{\beta _{p}}}$ & $\alpha _{q}$ \\ \hline
$e_{M}(\rho _{\alpha })-c$ & $\frac{1}{2}\log \left[ \left( \frac{1}{4\alpha
_{p}}+\beta _{q}\right) \right. $ & $\log \left( \sqrt{\beta _{q}\beta _{p}}%
+1/2\right) $ & $\frac{1}{2}\log \left[ \left( \frac{1}{4\alpha _{q}}+\beta
_{p}\right) \right. $ \\
& $\times \left( \alpha _{p}+\beta _{p}\right) ]$ &  & $\times \left( \alpha
_{q}+\beta _{q}\right) ]$ \\ \hline
$C(M;\alpha )$ & $\frac{1}{2}\log \frac{\alpha _{q}+\beta _{q}}{\frac{1}{%
4\alpha _{p}}+\beta _{q}}$ & $\frac{1}{2}\log \frac{\left( \alpha _{q}+\beta
_{q}\right) \left( \alpha _{p}+\beta _{p}\right) }{\left( \sqrt{\beta
_{q}\beta _{p}}+1/2\right) ^{2}}$ & $\frac{1}{2}\log \frac{\alpha _{p}+\beta
_{p}}{\frac{1}{4\alpha _{q}}+\beta _{p}}$ \\ \hline
\end{tabular}%
\end{equation*}%
Here the column C corresponds to the case where the \textquotedblleft
threshold condition\textquotedblright\ holds, implying (\ref{e=min}). Then
the full validity of the HGM in much more general multimode situation was
established in \cite{acc-noJ}. All the quantities in this column as well as
the value of $C(M,H,E)$ in the central column of the table 2 were obtained
in that paper as an example. On the other hand, the HGM remains open in the
cases of mutually symmetric columns L and R (for the derivation of the
quantities in column L of tables 1, 2 see Appendix).

Maximizing $C(M;\alpha )$ over $\alpha _{q},\alpha _{p}$ which satisfy the
energy constraint (\ref{E11}) (with the equality): $\alpha _{q}+\alpha
_{q}=2E$, we obtain $C(M,H,E)$ depending on the signal energy $E$ and the
measurement noise variances $\beta _{q},\beta _{p}:$
\begin{equation*}
\begin{tabular}{|c|c|c|}
\hline
\multicolumn{3}{|c|}{$\text{Table 2: }C(M,H,E)$} \\ \hline
L: HGM & C: \cite{acc-noJ} & R: HGM \\ \hline
$\beta _{q}\leq \beta _{p};E<E\left( \beta _{p},\beta _{q}\right) $ & $E\geq
E\left( \beta _{p},\beta _{q}\right) \vee E\left( \beta _{q},\beta
_{p}\right) $ & $\beta _{p}\leq \beta _{q};E<E\left( \beta _{q},\beta
_{p}\right) $ \\ \hline
$\log \left( \frac{\sqrt{1+8E\beta _{q}+4\beta _{q}^{2}}-1}{2\beta _{q}}%
\right) $ & $\log \left( \frac{E+\left( \beta _{q}+\beta _{p}\right) /2}{%
\sqrt{\beta _{q}\beta _{p}}+1/2}\right) $ & $\log \left( \frac{\sqrt{%
1+8E\beta _{p}+4\beta _{p}^{2}}-1}{2\beta _{p}}\right) $ \\ \hline
\end{tabular}%
\end{equation*}%
where we introduced the \textquotedblleft energy threshold
function\textquotedblright\
\begin{equation*}
E\left( \beta _{1},\beta _{2}\right) =\frac{1}{2}\left( \beta _{1}-\beta
_{2}+\sqrt{\frac{\beta _{1}}{\beta _{2}}}\right) .
\end{equation*}

Let us stress that, opposite to column C, the values of $C(M,H,E)$ in the L
and R columns are hypothetic, conditional upon validity of the HGM. Looking
into the left column, one can see that $C(M;\alpha )$ and $C(M,H,E)$ do not
depend at all on $\beta _{p}.$ Thus we can let $\beta _{p}\rightarrow
+\infty ,$ and in fact set $\beta _{p}=+\infty ,$ which corresponds to the
approximate measurement of position $q$ with Gaussian noise described by
POVM
\begin{equation}
M(dx)=\exp \left[ -\frac{\left( q-x\right) ^{2}}{2\beta _{q}}\right] \frac{dx%
}{\sqrt{2\pi \beta _{q}}}=D(x,0)\mathrm{e}^{-q^{2}/2\beta _{q}}D(x,0)^{\ast }%
\frac{dx}{\sqrt{2\pi \beta _{q}}},  \label{apprq}
\end{equation}%
which belongs to type 2 according to the classification of \cite{hclass}. In
other words, one makes the \textquotedblleft classical\textquotedblright\
measurement of the observable%
\begin{equation*}
X=q+\xi ,\quad \xi \sim \mathcal{N}(0,\beta _{q}),
\end{equation*}%
with the quantum energy constraint $\mathrm{Tr}\,\rho (q^{2}+p^{2})\leq 2E$.

The measurement channel corresponding to POVM (\ref{apprq}) acts on the
centered Gaussian state $\rho _{\alpha }\ $ by the formula
\begin{equation}
M:\rho _{\alpha }\rightarrow p_{\rho _{\alpha }}(x)=\frac{1}{\sqrt{2\pi
\left( \alpha _{q}+\beta _{q}\right) }}\exp \left[ -\frac{x^{2}}{2\left(
\alpha _{q}+\beta _{q}\right) }\right] .  \label{1}
\end{equation}%
In this case we have
\begin{eqnarray}
h_{M}(\rho _{\alpha }) &=&\frac{1}{2}\log \left( \alpha _{q}+\beta
_{q}\right) +c,  \label{ha2} \\
e_{M}(\rho _{\alpha }) &=&\frac{1}{2}\log \left( 1/\left( 4\alpha
_{p}\right) +\beta _{q}\right) +c,  \label{em2}
\end{eqnarray}%
which differ from the values in the case of finite $\beta _{p}\rightarrow
+\infty $ by the absence of the factor $\left( \alpha _{p}+\beta _{p}\right)
$ under the logarithms, while the difference $C(M;\alpha )=h_{M}(\rho
_{\alpha })-e_{M}(\rho _{\alpha })$ and the capacity $C(M,H,E)$ have the
same expressions as in that case (column L).

For $\beta _{q}=0$ (sharp position measurement, type 3 of \cite{hclass}) the
HGM is valid with
\begin{equation*}
C(M,H,E)=\log 2E.
\end{equation*}%
This follows from the general upper bound
\begin{equation}
C(M,H,E)\leq \log \left( 1+\frac{E-1/2}{\beta _{q}+1/2}\right) =\log \left(
\frac{2(E+\beta _{q})}{1+2\beta _{q}}\right)  \label{upg}
\end{equation}%
for $\beta _{q}\geq 0$ (Eq. (28) in \cite{hall2}, see also Eq. (5.39) in
\cite{caves}).

\section{The dual problem: accessible information}

Let us sketch here \textit{ensemble-observable duality} \cite{hall}, \cite%
{da}, \cite{h5} (see \cite{acc} for detail of mathematically rigorous
description in the infinite dimensional case).

Let $\mathcal{E}=\left\{ \pi (dx),\rho (x)\right\} $ be an ensemble, $\mu
(dy)$ a $\sigma -$finite measure and $M=\left\{ M(dy)\right\} $ an
observable having operator density $m(y)=M(dy)/\mu (dy)$ with values in the
algebra of bounded operators in $\mathcal{H}$. The dual pair
ensemble-observable $\left\{ \mathcal{E}^{\prime },M^{\prime }\right\} $ is
defined by the relations
\begin{equation}
\mathcal{E}^{\prime }:\quad \pi ^{\prime }(dy)=\mathop{\rm Tr}\nolimits\bar{%
\rho}_{\mathcal{E}}\,M(dy),\quad \rho ^{\prime }(y)=\frac{\bar{\rho}_{%
\mathcal{E}}^{1/2}m(y)\bar{\rho}_{\mathcal{E}}^{1/2}}{\mathop{\rm Tr}%
\nolimits\bar{\rho}_{\mathcal{E}}\,m(y)};  \label{piprime}
\end{equation}%
\begin{equation}
M^{\prime }:\quad M^{\prime }(dx)=\bar{\rho}_{\mathcal{E}}^{-1/2}\rho (x)%
\bar{\rho}_{\mathcal{E}}^{-1/2}\pi (dx),  \label{mprime}
\end{equation}%
Then the average states of both ensembles coincide
\begin{equation}
\bar{\rho}_{\mathcal{E}}=\bar{\rho}_{\mathcal{E}^{\prime }}  \label{III}
\end{equation}%
and the joint distribution of $x,y$ is the same for both pairs $(\mathcal{E}%
,M)$ and $(\mathcal{E}^{\prime },M^{\prime })$ so that%
\begin{equation}
I(\mathcal{E},M)=I(\mathcal{E}^{\prime },M^{\prime }).  \label{II}
\end{equation}%
Moreover,
\begin{equation}
\sup_{M}I(\mathcal{E},M)=\sup_{\mathcal{E}^{\prime }:\bar{\rho}_{\mathcal{E}%
^{\prime }}=\bar{\rho}_{\mathcal{E}}}I(\mathcal{E}^{\prime },M^{\prime }),
\label{inf}
\end{equation}%
where the supremum in the right-hand side is taken over all ensembles $%
\mathcal{E}^{\prime }$ satisfying the condition $\bar{\rho}_{\mathcal{E}%
^{\prime }}=\bar{\rho}_{\mathcal{E}}$. It can be shown (\cite{acc},
Proposition 4), that the supremum in the lefthand side remains the same if
it is taken over \textit{all} observables $M$ (not only of the special kind
with the density we started with), and then it is called the \textit{%
accessible information} $A(\mathcal{E)}$ of the ensemble $\mathcal{E}$. Thus
\begin{equation*}
A(\mathcal{E})=\sup_{\mathcal{E}^{\prime }:\bar{\rho}_{\mathcal{E}^{\prime
}}=\bar{\rho}_{\mathcal{E}}}I(\mathcal{E}^{\prime },M^{\prime }).
\end{equation*}%
Since the application of the duality to the pair $\left\{ \mathcal{E}%
^{\prime },M^{\prime }\right\} $ results in the initial pair $\left\{
\mathcal{E},M\right\} ,$ we also have
\begin{equation*}
A(\mathcal{E}^{\prime })=\sup_{M^{\prime }}I(\mathcal{E}^{\prime },M^{\prime
})=\sup_{\mathcal{E}:\bar{\rho}_{\mathcal{E}}=\bar{\rho}_{\mathcal{E}%
^{\prime }}}I(\mathcal{E},M).
\end{equation*}

Coming to the case of bosonic mode, we fix the Gaussian state $\rho _{\alpha
}$ and restrict to ensembles $\mathcal{E}$ with $\bar{\rho}_{\mathcal{E}%
}=\rho _{\alpha }.$ Let $M$ be the measurement channel corresponding to POVM
(\ref{MTB}). Then according to formulas (\ref{piprime}), the dual ensemble $%
\mathcal{E}^{\prime }=\left\{ p^{\prime }(x,y),\,\rho ^{\prime
}(x,y)\right\} ,$ where $p^{\prime }(x,y)$ is the Gaussian probability
density (\ref{pdmtb}) and
\begin{equation*}
\,\rho ^{\prime }(x,y)=\left[ p^{\prime }(x,y)\right] ^{-1}\sqrt{\rho
_{\alpha }}D(x,y)\rho _{\beta }D(x,y)^{\ast }\sqrt{\rho _{\alpha }}.
\end{equation*}%
By using the formula for $\sqrt{\rho_1}\rho_2\sqrt{\rho_1}$ where $\rho_1,
\rho_2$ are Gaussian operators (see \cite{lami} and also Corollary in the
Appendix of \cite{acc2}), we obtain%
\begin{equation*}
\rho ^{\prime }(x,y)=D(x^{\prime },y^{\prime })\rho _{\alpha ^{\prime
}}D(x^{\prime },y^{\prime })^{\ast }=\rho _{\alpha ^{\prime }}(x^{\prime
},y^{\prime }),
\end{equation*}%
where
\begin{equation}
\alpha ^{\prime }=\alpha -\gamma ^{\prime },\quad \gamma ^{\prime }=\kappa
\left( \alpha +\beta \right) ^{-1}\kappa ,\quad \left[
\begin{array}{c}
x^{\prime } \\
y^{\prime }%
\end{array}%
\right] =\kappa \left( \alpha +\beta \right) ^{-1}\left[
\begin{array}{c}
x \\
y%
\end{array}%
\right] .  \label{prime}
\end{equation}%
and%
\begin{equation}
\kappa =\sqrt{I+\left( 2\alpha \Delta ^{-1}\right) ^{-2}}\,\alpha =\alpha
\sqrt{I+\left( 2\Delta ^{-1}\alpha \right) ^{-2}}.  \label{kappa}
\end{equation}%
Since $\left[
\begin{array}{cc}
x & y%
\end{array}%
\right] ^{t}\sim\mathcal{N}(0,\alpha +\beta ),$ then from second and third
equations in (\ref{prime}) we obtain $\left[
\begin{array}{cc}
x^{\prime } & y^{\prime }%
\end{array}%
\right] ^{t}\sim\mathcal{N}(0,\kappa \left( \alpha +\beta \right)
^{-1}\kappa )=\mathcal{N}(0,\gamma ^{\prime }).$ By denoting $p_{\gamma
^{\prime }}(x^{\prime },y^{\prime })$ the density of this normal
distribution, we can equivalently rewrite the ensemble $\mathcal{E}^{\prime}$
as $\mathcal{E}^{\prime }=\left\{ p_{\gamma ^{\prime }}(x^{\prime
},y^{\prime }),\,\rho _{\alpha ^{\prime }}(x^{\prime },y^{\prime })\right\} $
with the average state $\rho _{\alpha },$ $\alpha =\alpha ^{\prime }+\gamma
^{\prime }.$ Then HGM is equivalent to the statement
\begin{equation*}
A(\mathcal{E}^{\prime })=C\left( M;\alpha \right) ,
\end{equation*}%
where the values of $C\left( M;\alpha \right) $ are given in the table 1,
however they should be reexpressed in terms of the ensemble parameters $%
\gamma ^{\prime },\alpha ^{\prime }$. In \cite{acc2} we treated the case C
in multimode situation, establishing that the optimal measurement is
Gaussian, and described it. Here we will discuss the case L (R is similar)
and show that for large $\beta _{p}$ (including $\beta _{p}=+\infty $) the
HGM is equivalent to the following: the value of the accessible information
\begin{equation*}
A(\mathcal{E}^{\prime })=C\left( M;\alpha \right) =\frac{1}{2}\log \frac{%
\alpha _{q}+\beta _{q}}{\frac{1}{4\alpha _{p}}+\beta _{q}}
\end{equation*}%
is attained on the sharp position measurement $M_{0}^{\prime
}(d\xi)=|\xi\rangle \langle \xi |d\xi$ (in fact this refers to the whole
domain L: $\frac{1}{2}\sqrt{\frac{\beta _{q}}{\beta _{p}}}<\frac{1}{4\alpha
_{p}},$ which however has rather cumbersome description in the new variables
$\gamma ^{\prime },\alpha ^{\prime }$, cf. \cite{acc2}).

In the one mode case we are considering the matrix $\alpha $ is given by (%
\ref{alpha}), $\beta $ -- by (\ref{beta}), and $\Delta =\left[
\begin{array}{cc}
0 & 1 \\
-1 & 0%
\end{array}%
\right] ,$ so that $\left( 2\Delta ^{-1}\alpha \right) ^{2}=-\left( 4\alpha
_{q}\alpha _{p}\right) I.$ Computations according to (\ref{prime}) and (\ref%
{kappa}) give
\begin{equation}
\alpha ^{\prime }=\left[
\begin{array}{cc}
\alpha _{q}^{\prime } & 0 \\
0 & \alpha _{p}^{\prime }%
\end{array}%
\right] =\left[
\begin{array}{cc}
\frac{\alpha _{q}\left( \beta _{q}+1/\left( 4\alpha _{p}\right) \right) }{%
\alpha _{q}+\beta _{q}} & 0 \\
0 & \frac{\alpha _{p}\left( \beta _{p}+1/\left( 4\alpha _{q}\right) \right)
}{\alpha _{p}+\beta _{p}}%
\end{array}%
\right] .  \label{aptime}
\end{equation}%
But under the sharp position measurement $M_{0}^{\prime }(d\xi )=|\xi
\rangle \langle \xi |d\xi ,$ one has \footnote{%
In the formulas below $p(\xi)=\mathcal{N}(m,\alpha )$ means that $p(\xi)$ is
Gaussian probability density with mean $m$ and variance $\alpha$.}

\begin{equation*}
p(\xi |x^{\prime },y^{\prime })=\langle \xi |\,\rho _{\alpha ^{\prime
}}(x^{\prime },y^{\prime })|\xi \rangle =\mathcal{N}(x^{\prime },\alpha
_{q}^{\prime })
\end{equation*}%
while $\langle \xi |\,\rho _{\alpha }|\xi \rangle =\mathcal{N}(0,\alpha _{q})
$ (note that $\bar{\rho}_{\mathcal{E}^{\prime }}=\bar{\rho}_{\mathcal{E}%
}=\rho _{\alpha }$) and%
\begin{eqnarray}
I\left( \mathcal{E}^{\prime },M_{0}^{\prime }\right)  &=&\frac{1}{2}\left[
\log \left( \alpha _{q}^{\prime }+\gamma _{q}^{\prime }\right) -\log \alpha
_{q}^{\prime }\right]   \notag \\
&=&\frac{1}{2}\left[ \log \alpha _{q}-\log \frac{\alpha _{q}\left( \beta
_{q}+1/4\alpha _{p}\right) }{\left( \alpha _{q}+\beta _{q}\right) }\right]
\notag \\
&=&\frac{1}{2}\log \frac{\left( \alpha _{q}+\beta _{q}\right) }{\left( \beta
_{q}+1/4\alpha _{p}\right) },  \label{I3}
\end{eqnarray}%
which is identical to the expression in (\ref{cg2}).

In the case of the position measurement channel $M$ corresponding to POVM (%
\ref{apprq}) ($\beta _{p}=+\infty )$ we have $\alpha _{p}^{\prime }=\alpha
_{p},$ otherwise the argument is essentially the same. Thus we obtain that
the HGM concerning $e_{M}(\rho )$ in the case L is equivalent to the
following:

\textit{The accessible information of a Gaussian ensemble \ $\mathcal{E}%
^{\prime }=\left\{ p^{\prime }(x),\,\rho ^{\prime }(x)\right\} ,$ where%
\begin{equation*}
p^{\prime }(x)=\mathcal{N}(0,\gamma _{q}^{\prime }),\quad \rho ^{\prime
}(x)=D(x,0)\rho _{\alpha ^{\prime }}D(x,0)^{\ast },\quad
\end{equation*}%
is given by the expression (\ref{I3}) and attained on the sharp position
measurement $M_{0}^{\prime }(dx)=|\xi \rangle \langle \xi |d\xi .$}

\section{Appendix. Case L in tables 1, 2}

\label{app}

By taking the Gaussian ensemble parameters in (\ref{em}) as
\begin{equation}
\delta =1/\left( 4\alpha _{p}\right) ,\quad \gamma _{p}=0,\quad \gamma
_{q}=\alpha _{q}-1/\left( 4\alpha _{p}\right) ,  \label{param}
\end{equation}%
we get the hypothetic value
\begin{equation}
e_{M}(\rho _{\alpha })=\frac{1}{2}\log \left( \frac{1}{4\alpha _{p}}+\beta
_{q}\right) \left( \alpha _{p}+\beta _{p}\right)+c ,  \label{em1}
\end{equation}%
hence taking into account (\ref{ha}),
\begin{equation}
C(M;\alpha )=h_{M}(\rho _{\alpha })-e_{M}(\rho _{\alpha })=\frac{1}{2}\log
\frac{\alpha _{q}+\beta _{q}}{\frac{1}{4\alpha _{p}}+\beta _{q}}.
\label{cg2}
\end{equation}%
The constrained capacity is
\begin{eqnarray}
C(M,H,E) &=&\max_{\alpha _{q}+\alpha _{q}\leq 2E}\frac{1}{2}\left[ \log
\left( \alpha _{q}+\beta _{q}\right) -\log \left( 1/\left( 4\alpha
_{p}\right) +\beta _{q}\right) \right]  \label{cg1} \\
&=&\max_{\alpha _{p}}\frac{1}{2}\left[ \log \left( 2E-\alpha _{p}+\beta
_{q}\right) -\log \left( 1/\left( 4\alpha _{p}\right) +\beta _{q}\right) %
\right] ,  \notag
\end{eqnarray}%
where in the second line we took the maximal value $\alpha _{q}=2E-\alpha
_{p}$. Differentiating, we obtain the equation for the optimal value $\alpha
_{p}$:
\begin{equation*}
4\beta _{q}\alpha _{p}^{2}+2\alpha _{p}-\left( 2E+\beta _{q}\right) =0,
\end{equation*}%
the positive solution of which is%
\begin{equation}
\alpha _{p}=\frac{1}{4\beta _{q}}\left( \sqrt{1+8E\beta _{q}+4\beta _{q}^{2}}%
-1\right) ,  \label{ap}
\end{equation}%
whence%
\begin{equation}
C(M,H,E)=\log \left( \frac{\sqrt{1+8E\beta _{q}+4\beta _{q}^{2}}-1}{2\beta
_{q}}\right) .  \label{capg}
\end{equation}%
The parameters of the optimal Gaussian ensemble are obtained by substituting
the value (\ref{ap}) into (\ref{param}) with $\alpha _{q}=2E-\alpha _{p}$.
\bigskip

The above derivation concerns the measurement (\ref{MTB}) ($\beta
_{p}<\infty ).$ The case of the measurement (\ref{apprq}) ($\beta
_{p}=+\infty )$ is treated similarly, with (\ref{em1}), (\ref{ha}) replaced
by (\ref{em2}), (\ref{ha2}). Notably, in this case the expression (\ref{capg}%
) coincides with the one obtained in \cite{hall3} by optimizing the
information from applying sharp position measurement to noisy optimally
squeezed states\footnote{%
The author is indebted to M. J. W. Hall for this observation.}. \bigskip

\textbf{Acknowledgment}. 
The author is grateful to M. J. W. Hall for sending a copy of his paper \cite%
{hall3}, and to M. E. Shirokov for the comments improving the presentation.


\begin{thebibliography}{99}
\bibitem{sera} Serafini A., \emph{Quantum Continuous Variables: A Primer of
Theoretical Methods}, CRC Press, Taylor \& Francis Group, 2017.


\bibitem{caves} Caves C.M., Drummond P.D. Quantum limits on bosonic
communication rates. Rev. Mod. Phys. 1994, vol. 68, N2, 481-537.

\bibitem{hall} Hall M. J. W., Quantum information and correlation bounds,
\emph{Phys. Rev. A} vol. 55, pp. 1050-2947, 1997.


\bibitem{hall2} Hall M. J. W., Information exclusion principle for
complementary observables, Phys. Rev. Lett. 74, 3307, 1995.

\bibitem{hall3} Hall M. J. W., Gaussian noise and quantum optical
communication, \emph{Phys. Rev. A} vol. 50, pp. 3295-3303, 1994.

\bibitem{QSCI} Holevo A. S., \emph{Quantum systems, channels, information: a
mathematical introduction}, 2-nd ed., Berlin/Boston: De Gruyter, 2019.

\bibitem{h5} Holevo A. S., Information capacity of quantum observable,
Problems Inform. Transmission, \textbf{48}:1, 1--10 (2012). arXiv:1103.2615.

\bibitem{h2} Holevo A. S., On the constrained classical capacity of
infinite-dimensional covariant channels \textit{J. Math. Phys.} \textbf{57}%
:1 15203 (2016).

\bibitem{acc} Holevo A. S., Gaussian maximizers for quantum Gaussian
observables and ensembles, IEEE Trans. Inform. Theory, 2020,
doi:10.1109/TIT.2020.2987789. 

\bibitem{ghm} Giovannetti V., Holevo A. S., Mari A., Majorization and
additivity for multimode bosonic Gaussian channels, Theor. Math. Phys.,
\textbf{182}:2, 284--293, (2015). arXiv:1405.4066

\bibitem{acc-noJ} Holevo A. S., Kuznetsova A. A., Information capacity of
continuous variable measurement channel. J. Phys. A: Math. Theor. \textbf{53}
(2020) 175304 (13pp.). 

\bibitem{hy} Holevo A. S., Yashin V. I., Quantum information aspects of
approximate position measurement, arXiv:2006.04383.

\bibitem{acc2} Holevo A. S., Accessible information of a general quantum
Gaussian ensemble, arXiv:2102.01981.

\bibitem{hclass} Holevo A. S., The structure of general quantum Gaussian
observable, arXiv:2007.02340.



\bibitem{Shir} Shirokov M. E., On entropic quantities related to the
classical capacity of infinite dimensional quantum channels, Theory of
Probability and its Applications, Vol. 52, No. 2, (2007), 250-276.
arXiv:quant-ph/0411091

\bibitem{Shir1} Shirokov M. E., On properties of the space of quantum states
and their application to the construction of entanglement monotones, Izv.
Math., \textbf{74}:4 (2010), 849-882.

\bibitem{wgc} Wolf M. M., Giedke G., Cirac J. I., Extremality of Gaussian
quantum states, Phys. Rev. Lett. \textbf{96}, 080502 (2006).%

\bibitem{huds} Cushen C. D., Hudson R. L., A quantum mechanical central
limit theorem. J. Appl. Prob. \textbf{8}, (1971) 454-469.

\bibitem{lami} Lami L., Das S., Wilde M. M., Approximate reversal of quantum
Gaussian dynamics, J. Phys. A, \textbf{51}:12, 125301, 2018.

\bibitem{da} Dall'Arno M., D'Ariano G. M., Sacchi M.F., Informational power
of quantum measurements, Phys. Rev. \textbf{A 83}, 062304 (2011).

\bibitem{Oreshkov} Oreshkov O.,Calsamiglia J., Munoz-Tapia R., Bagan E.,
Optimal signal states for quantum detectors, New J. Phys. \textbf{13}
(2011), 073032.



\bibitem{scha} Sch\"afer J., Karpov E., Gar\'cia-Patr\'on R., Pilyavets O.
V., Cerf N. J., Equivalence Relations for the Classical Capacity of
Single-Mode Gaussian Quantum Channels, Phys. Rev. Lett. \textbf{111}, (2013)
030503.


\bibitem{guha} Takeoka M., Guha S., Capacity of optical communication in
loss and noise with general Gaussian receivers, Phys. Rev. \textbf{A 89},
042309 (2014).

\bibitem{lee} Jaehak Lee, Se-Wan Ji, Jiyong Park, Hyunchul Nha, Gaussian
benchmark for optical communication aiming towards ultimate capacity, Phys.
Rev. \textbf{A 93}, 050302(R) (2016).


\end{thebibliography}
\end{document}